\documentstyle[preprint,revtex]{aps}
\begin{document}
\draft
\preprint{LSUHE No. 183-1994 }
\def\overlay#1#2{\setbox0=\hbox{#1}\setbox1=\hbox to \wd0{\hss #2\hss}#1%
\hskip -2\wd0\copy1}
\begin{title}
 Disappearance of the Abrikosov vortex above the deconfining\\
phase transition in SU(2) lattice gauge theory
\end{title}
\author{ Yingcai Peng and Richard W. Haymaker}
\begin{instit}
Department of Physics and Astronomy, \\
Louisiana State University, Baton Rouge, Louisiana 70803-4001
\end{instit}
\begin{abstract}
       We calculate the solenoidal magnetic monopole current and
electric flux distributions at finite temperature in the presence of
a static quark-antiquark pair. The simulation was performed using
SU(2) lattice gauge theory in the maximal Abelian gauge.
We find that the monopole current and electric flux distributions
are quite different below and above the finite temperature
deconfining phase transition point and agree with predictions of the
Ginzburg-Landau theory.

\end{abstract}

\narrowtext
\section{Introduction}
\label{sec: intro}

   It is widely accepted that quark confinement is due to
color-electric flux tube formation. In this scenario the QCD
vacuum behaves like a dual superconductor.  That is, magnetic monopoles
in the QCD vacuum respond to a color-electric field by generating screening
currents to confine the electric flux into a narrow tube dual to
the Abrikosov vortex arising from the Meissner effect.
Some direct evidence for this scenario has
been reported for U(1) LGT and SU(N) lattice gauge theories (LGT)
simulations~\cite{sbh1,sbh,sbh2,hsbw,mes}.
 It is found that monopoles in U(1) and
the $U(1)^{N-1}$ abelian projection of SU(N)
LGT in the maximal Abelian gauge\cite{th} react to the
electric flux from a static $q\bar q$
pair by producing a solenoidal current distribution which confines
the flux to a narrow tube.

It is clear that the solenoidal magnetic monopole
currents are associated with the confined flux tube. At
finite temperatures the flux tube is expected to exist only in the
low-temperature confined phase, but not in the high-temperature
unconfined phase. Previous studies~\cite{sbh1,sbh,sbh2,hsbw,mes}
 addressed the issues of
monopole current distributions and the structures of the electric flux
tube at zero temperature.
Questions remain as to the monopole current distributions at
finite temperature, especially in the unconfined phase.
In this paper we shall present our observations of monopole current
and electric flux distributions at finite temperature. This study is
a direct extension of the earlier work~\cite{sbh,sbh2} in SU(2) LGT.

\bigskip
\section{Simulations}
\label{sec: thb}

    Our simulations were performed on a lattice of the size
$4 \times 17^2 \times 19$ with skew-periodic boundary conditions.
Each link from point $s$ in the $\mu$ direction carries an $SU(2)$
element $U(s, \mu)$, and the plaquette operator $U_P(s)$ is formed
in the usual fashion as a directed product of link  variables,
where the label $P$ represents the orientation of the plaquette, $(\mu, \nu)$.
The standard Wilson action~\cite{cr} is used in our calculations
\begin{equation}
 S(U)=\beta \sum_{s,P}(1-\frac{1}{2} {\rm Tr} U_P).
\label{e2.1}
\end {equation}
We generated lattice configurations with the distribution $\exp (-S)$
using a combination of the standard Metropolis algorithm~\cite{met} and
overrelaxation~\cite{cr1}. Simulations were performed for
$\beta=2.25$, 2.28 and 2.40.  In each case the initial 3000 sweeps were used
to thermalize the system, then one measurement was made every
50 sweeps.

   Each measurement was performed on a lattice configuration
in the maximal Abelian gauge~\cite{th}, which can be achived by
finding the gauge transformation that maximized the quantity~\cite{ksw}
\begin{equation}
 R= \sum_{s, \mu} {\rm Tr} [\sigma_3 U(s, \mu) \sigma_3 U^{\dagger}
         (s, \mu) ].
\label{e2.2}
\end {equation}
This is equivalent to diagonalizing
\begin{equation}
 X(s)= \sum_{\mu} [ U(s, \mu) \sigma_3 U^{\dagger}(s, \mu) +
U^{\dagger}(s-\mu, \mu) \sigma_3 U(s-\mu, \mu) ].
\label{e2.3}
\end {equation}
at each site $s$. To measure the efficacy of gauge
fixing in the simulations
we used the lattice sum of the magnitude of the off-diagonal
component of $X(s)$, $|Z|^2=\sum_s |X(s)_{12}|^2$. Typically we
needed about 620 gauge fixing sweeps to attain
$|Z|^2\approx 10^{-5}$/site for each $\beta$ value.
Three gauge fixing methods were used: (1) generating and
accepting random local changes only if $R$ increased, (2) locally
maximizing $R$ exactly at alternate sites, (3) applying overrelaxation
using the square of the gauge transformation of method (2) to sample
configurations better.

\section{ Measurements }
\label{sec: mea}

   After gauge fixing, the Abelian $U(1)$ link variable is
given~\cite{ksw} by the phase of the diagonal component of the
$SU(2)$ link variable, $u(s, \mu)=U(s, \mu)_{11}/ |U(s, \mu)_{11}|$.
We construct the Abelian plaquette variable $p_{\mu \nu}(s)$
from $u(s, \mu)$ in the usual fashion. In this study we use
Abelian Polyakov loops to replace Abelian Wilson loop used in
ref.~\cite{sbh,sbh2}, in order to study the finite temperature case.
The Abelian Polyakov loop closed in the time direction can be
contructed as
\begin{equation}
 P_{ab}(s)=\prod_{\tau =1}^{N_t}u(s,\tau),
\label{e3.1}
\end {equation}
where $N_t$ is the temporal size of the lattice,
 $N_t\times N_s^2\times N_z$, specifically $N_t=4$ for the
lattice $4 \times 17^2 \times 19$.

    The magnetic monopoles are identified using the
DeGrand-Toussaint construction~\cite{dt}. For each Abelian
plaquette $p_{\mu \nu}(s)$ we decompose its phase angle
$\theta_p = arg ( p_{\mu \nu}(s) )$ ($ -4\pi \le \theta_p \le 4\pi$)
into two parts,
\begin{equation}
 \theta_p= {\bar \theta}_p + 2\pi n_p,
\label{e3.2}
\end {equation}
with $ \pi < {\bar \theta}_p < \pi$, where $n_p$ describes the
number of Dirac strings and ${\bar \theta}_p$ the electromagnetic
flux through the plaquette~\cite{dt}. The magnetic monopole current
density can be constructed as~\cite{hkk},
\begin{equation}
 K_\mu (s) = \epsilon_{\mu \nu \rho \sigma}
  [ {\bar \theta}_p(s+ \nu, \rho \sigma) -
    {\bar \theta}_p(s, \rho \sigma) ]/4\pi .
\label{e3.3}
\end {equation}
It is convenient to associate the monopole current density $K_\mu (s)$
in each three-volume with a link on the dual lattice~\cite{sbh}, making
world lines that define a conserved current density ${\bf J_M}$.
To isolate the solenoidal monopole currents we construct the operator for
the line integral of ${\bf J_M}$ around a dual plaquette,
${\bf \nabla \times J_M }$, as described in ref.~\cite{sbh}.

    In our measurements we chose two Abelian Polyakov loops
correlated along the $N_z$ direction, $\langle P_{ab}(0)
P^{\dagger}_{ab}(R) \rangle$, with the separation $R=3a$,
$4a$, $5a$ and $6a$, which represent the static $q \bar q$ pair.
The electric field operator is given by $a^2 {\cal E}_i = {\rm Im} p_{i4}$,
where $a$ is the lattice spacing~\cite{sbh}.
Then the electric field in the $N_z$ direction can be calculated as,
\begin{equation}
 \langle E_z(R, {\bf x}) \rangle = \frac {\langle P_{ab}(0) P^{\dagger}_{ab}(R)
     {\rm Im}(p_{zt} ({\bf x}) )  \rangle }
      { a^2 e \langle P_{ab}(0) P^{\dagger}_{ab}(R) \rangle  },
\label{e3.4}
\end {equation}
with $e$ the Abelian electric charge and $p_{zt}$ the Abelian plaquette
in the $z-t$ plane.
The solenoidal monopole current distribution can be calculated
from the correlation~\cite{sbh},
\begin{equation}
 \langle \nabla \times J_M (R, {\bf x}) \rangle _z =
 \frac { 2\pi \langle P_{ab}(0)
 P^{\dagger}_{ab}(R) (\nabla \times J_M ({\bf x}) )_z   \rangle }
      { a^4 e \langle P_{ab}(0) P^{\dagger}_{ab}(R) \rangle  },
\label{e3.5}
\end {equation}
where the solenoidal monopole current operator, $(\nabla \times J_M )_z$,
is constructed as the dual plaquette in the $x-y$ plane.
In practice, the expressions for $\langle E_z(R, {\bf x}) \rangle$
and $\langle \nabla \times J_M( R, {\bf x}) \rangle _z$ can be simplified
further. If we denote $\theta_{PP^{\dagger}}$ as the phase angle of the
product of two Abelian Polyakov loops, $( P_{ab}(0)P^{\dagger}_{ab}(R) )$,
Eq.~(\ref{e3.4}) can be written as,
\begin{equation}
 \langle E_z(R, {\bf x}) \rangle = \frac {\langle sin \theta_{PP^{\dagger}}
     sin \theta_p \rangle }
      { a^2 e \langle cos \theta_{PP^{\dagger}} \rangle  } ,
\label{e3.6}
\end {equation}
where $\theta_p$ is the phase angle of the Abelian plaquette $p_{zt}$.
Also, Eq.~(\ref{e3.5}) becomes
\begin{equation}
 \langle \nabla \times J_M( R, {\bf x}) \rangle _z = \frac
       { 2\pi \langle sin \theta_{PP^{\dagger}}
         (\nabla \times J_M ({\bf x}) )_z \rangle }
      { a^4 e \langle cos \theta_{PP^{\dagger}} \rangle  } .
\label{e3.7}
\end {equation}

\section{ Numerical Results }
\label{sec: nur}

     As we choose the lattice extent $N_t$ of the 4-dimensional lattice,
$N_t \times N_s^2 \times N_z$, as the time direction, we can
simulate the finite temperature case on the lattice with the
temperature $T=1/N_t a(\beta)$, where the lattice spacing $a$ is a
function of $\beta$. For SU(2) lattice gauge theory the finite
temperature deconfining phase transition point has been studied
extensively. One recent simulation result~\cite{efm} is,
for $N_t=4$ the transition point $\beta_c$ is given as,
\begin{equation}
\beta_c = 2.2986\pm 0.0006.
\label{e4.1}
\end {equation}
Below $\beta_c$ ($\beta < \beta_c$) the system is in the confined phase,
but above $\beta_c$ it is in the unconfined phase.

   We measured the electric flux $\langle E_z(R, {\bf x}) \rangle$
and the solenoidal monopole current
$\langle \nabla \times J_M( R, {\bf x}) \rangle_z$ on the transverse
plane midway between the $q \bar q$ pair for several $\beta$
values, $\beta=2.25$, 2.28 and 2.40. Two $\beta$ values
(2.25 and 2.28) are below the transition point $\beta_c$ given
by Eq.~(\ref{e4.1}), one $\beta$ value (2.40) is above the $\beta_c$.
In the following we proceed to discuss our results in the two cases,
which correspond to the confined and the unconfined phases separately.

\subsection{ Results in the confined phase }
\label{subsec: con}

     We measured the electric flux and the solenoidal monopole current
distributions of a $q \bar q$ pair in the confined
phase: $\beta =2.25$ and 2.28,  In each case we accumulated about 800
measurements.

     Fig.~\ref{f4.1} shows the electric field $E_z$ distribution
for $\beta =2.28$, with the $q \bar q$ separation $R=3a$, $4a$, $5a$
and $6a$ separately. This displays the strength of $E_z$ versus the
distance $r$ from the $q \bar q$ axis on the transverse plane midway
between the $q \bar q$ pair. For small quark separations, e.g.
$R=3a$ and $4a$, as shown in Fig.~\ref{f4.1}(a) and (b), our
results are consistent with those of the earlier works, which
were obtained from simulations using Abelian Wilson loops of sizes
$3 \times 3$ in ref.\cite{sbh} and $3 \times 3$ and $5 \times 5$ in
ref.\cite{mes}.   In this study we observe
clear signals for $E_z$ at $R = 5a,6a$ as shown in Fig.~\ref{f4.1}(c)
and (d). From Fig.~\ref{f4.1} one can see that the electric flux $E_z$
has the peak value on the $q \bar q$ axis ($r=0$), then it decreases
with the off-axis distance $r$ rapidly. As the $q \bar q$ separation $R$
increases the $E_z$ values also decrease as we expected. Our data
also shows that the electric flux $E_z$ basically vanishes
within errors at the off-axis distance $r\approx 5a$ for all
$R$ values in Fig.~\ref{f4.1}(a)-(d). This agrees with the expectation
that electric flux is squeezed into a narrow tube in the confined phase.

     In Fig.~\ref{f4.2} we show the monopole current
$-(\nabla \times J_M)_z$ as a function of the off-axis distance $r$
for $\beta =2.28$, with $R=3a$ and $4a$. Again our data shows
that the curl of the monopol current $-(\nabla \times J_M)_z$
has the same functional form as those of the earlier works
for SU(2) LGT~\cite{sbh,sbh2}.  Fig.~\ref{f4.2} only displays
the results at small quark separations, e.g. $R=3a$ and $4a$ since
for large quark separations, e.g. $R=5a$ and $6a$
the noise is too large. The large signals shown in Fig.~\ref{f4.2}
confirm that the solenoidal monopole
current surrounding the electric flux tube is large in the
finite temperature confined phase.
For $\beta =2.25$ the results are similar to the case of
$\beta =2.28$.

To clarify the significance of these profiles, we wish to summarize
briefly the results of ref.\cite{sbh1,sbh}.  Picture for the moment
the points in Fig.~\ref{f4.2} at $r/a = 1, 1.4$ to be negative and
further that the curve matchs, up to a factor, the negative of the electric
field profile for $r/a \neq 0$.  This is precisely what we found for
U(1), ref.\cite{sbh1}, and is the prediction of the London
theory.  The linear combination of $E_z - \lambda^2 (\nabla \times J_M)_z$
is defined as the fluxoid, where $\lambda$ is the London penetration
depth.  For this geometry the fluxoid gives a lattice delta function at
the origin, with the coefficient equal to the total electric flux in the
vortex.

The departure of  Fig.~\ref{f4.2} from this picture is fully explained by
the more general Ginzburg-Landau theory in which the order parameter
for superconductivity turns on at a boundary over a finite distance
(called the coherence length) rather than abruptly at the boundary as in
the above case.   The coherence length in the Fig.~\ref{f4.2} is the
distance at which the profile deviates from the London theory which is
$r/a = \simeq 2.0$.   The London penetration depth is independently
determined by the exponential fall-off of $E_z$.

\subsection{ Results in the unconfined phase }
\label{subsec: uncon}

    We also studied the electric flux and monopole current distributions
in the unconfined phase. The data were measured at $\beta =2.40$, and
410 measurements were accumulated.

     Fig.~\ref{f4.3} shows the electric flux $E_z$ as a function of the
off-axis distance $r$ for $\beta =2.40$, with the $q \bar q$ separation
$R=3a$, $4a$, $5a$ and $6a$ separately. The significant difference between
Fig.~\ref{f4.3} and Fig.~\ref{f4.1} is that the $E_z$ flux values
for $\beta =2.40$ approach zero very slowly, even at large off-axis
distances (e.g. $r= 6a$) $E_z$ flux values still do not vanish
within errors, as shown clearly in Fig.~\ref{f4.3}(c) and (d).
However, Fig.~\ref{f4.1} shows that the $E_z$ flux data for $\beta =2.28$
decrease very quickly and vanish at $r\approx 5a$ within errors.
This implies that the electric flux spreads out from the $q \bar q$
axis in the unconfined phase ($\beta =2.40$), it is more Coulomb-like
than the flux distribution in the confined phase (e.g. $\beta =2.28$).

     Fig.~\ref{f4.4} shows the corresponding monopole current
distributions $-(\nabla \times J_M)_z$ for $\beta =2.40$,
with $R=3a$ and $4a$. From this figure one can see that almost
all data vanish within errors except one small value on the
$q \bar q$ axis ($r=0$) for each case. Comparing with the
monopole current distributions in the confined phase ($\beta =2.28$),
as shown in Fig.~\ref{f4.2} it is clear that the solenoidal monopol
current $-(\nabla \times J_M)_z $ becomes very small in
the unconfined phase, as expected.

\bigskip
\section{Conclusions}
\label{sec: sum}

    We extend the earlier studies~\cite{sbh,mes} on solenoidal monopole
current and electric flux distributions to the finite temperature
case. We find that both the electric flux and the monopole current have
quite different behaviors in the finite temperature confined and
unconfined phases. In the confined phase our data are consistent
with the results of earlier works~\cite{sbh,mes}, which correspond
to the zero-temperature case. We find that in the confined phase the
electric flux is confined in a narrow tube, and the solenoidal monopole
currents are large.
 However, in the unconfined phase our data show clearly that
the electric flux is not confined, and the solenoidal monopole
current is vanishingly small. This presents the significant
difference between the two phases.
This study confirms the point that the electric flux is repelled by
the surrounding solenoidal monopole currents, and is squeezed into
a narrow tube. In the unconfined phase
solenoidal monopole current vanishes, then the electric flux is
unconfined.

\bigskip
\nonum
\section{Acknowledgments}

  We wish to thank D. Browne, M. Polikarpov,
 H. Rothe,  G. Schierholz, V. Singh, T. Suzuki,
J. Wosiek and K. Yee
for many fruitful discussions on this problem.
We are especially grateful to V. Singh for making parts
of her code available to us.
 This research was supported by the U.S.
Department of Energy under Grant No. DE-FG05-91ER40617.

\figure{ Profile of the electric field $E_z$ as a function of the
transverse distance $r$ from the $q \bar q$ axis, with the
$q \bar q$ separation (a). $R= 3a$, (b) $R=4a$, (c) $R=5a$
and (d) $R=6a$. The data were
measured on the lattice $4 \times 17^2 \times 19$ with $\beta=2.28$.
The values of $E_z$ are measured in the lattice unit $1/a^2 e$. The
dashed lines are just to guide the eye.
\label{f4.1} }

\figure{ Profile of the solenoidal monopol current
$-(\nabla \times J_M)_z$ as a function of $r$, with the
$q \bar q$ separation (a). $R= 3a$ and (b) $R=4a$. The data were
measured on the lattice with $\beta=2.28$.
The values of $-(\nabla \times J_M)_z$ are measured in the lattice
unit $1/a^4 e$.
\label{f4.2} }

\figure{ Profile of $E_z$ vs. $r$ in the unconfined phase
with $\beta=2.40$. The $q \bar q$ separation is
 (a). $R= 3a$, (b) $R=4a$, (c) $R=5a$ and (d) $R=6a$.
The values of $E_z$ are measured in the lattice unit $1/a^2 e$.
\label{f4.3} }

\figure{ Profile of $(\nabla \times J_M)_z$ vs. $r$ in the unconfined
 phase with $\beta=2.40$. The $q \bar q$ separation is
 (a). $R= 3a$ and (b) $R=4a$.
The values of $-(\nabla \times J_M)_z$ are measured in the lattice
unit $1/a^4 e$.
\label{f4.4} }

\end{document}